\documentclass[prb,twocolumn,amssymb] {revtex4}
\usepackage{graphicx}
\begin{document}
\title{Emergent Geometric Hamiltonian and Insulator-Superfluid Phase Transitions}
\author{ Fei Zhou } 
\affiliation{ Department of Physics and
Astronomy, University of British Columbia,\\ 6224 Agricultural
Road, Vancouver, British Columbia, Canada, V6T 1Z1}
\date{\today}
\begin{abstract}
I argue that certain bosonic insulator-superfluid
phase transitions as an interaction constant varies are 
driven by emergent
geometric properties of insulating states. 
The {\em renormalized} chemical potential and population of disordered bosons at different levels
define the geometric aspect of an effective low energy Hamiltonian
which I employ to study various resonating states and quantum phase 
transitions. In a mean field approximation, I 
also demonstrate that the quantum phase transitions 
are in the universality class of a percolation problem.
\end{abstract}
\maketitle

\narrowtext

Localization or delocalization of interacting bosons 
has been a puzzling theoretical and experimental issue for quite long time
\cite{Anderson70,Reppy84,Ma86,Giamarchi88,Finotello88,Fisher89,Scalettar91,Krauth91,Altman04}.
Anderson {\em et al.} first investigated a possibility of bose condensation into an 
extended eigenstate
in disordered systems and pointed out certain instabilities of such condensates\cite{Anderson70}. 
Giamarchi and Schulz employed the replica-trick-based renormalization 
equations to study 
the quantum phase transitions between 
superfluid states and insulating states in one dimension
\cite{Giamarchi88}.
Fisher {\it et al.} on the other hand investigated the localization of 
interacting bosons in the context of a Bose-Hubbard 
(lattice) model\cite{Fisher89}.
In high-dimensions, this different approach appears to be very fruitful
resulting in various new observations; for instance a new distinct 
Bose-glass phase has been emphasized and its divergent superfluid 
susceptibility was pointed out. 
In the same context, numerical results further suggested 
additional Anderson-type insulating states in weak 
couplings\cite{Scalettar91}.
More recently, $1d$ superfluid-Mott-glass phase transitions in a 
particle-hole symmetric lattice
model have been studied\cite{Altman04}. 
Finally, a system of cold atoms in a random laser speckle potential has 
also been established and studied\cite{Lye04}.

However, little attention has been so far 
paid to the {\em microscopics} of bosonic insulators in the dilute limit.
Technically, the complexity of this issue is 
largely related to the failure of the 
replica-technique-based perturbative approach 
applied in this limit 
(in both 1D and high dimensions). 
This aspect of the problem seems likely to originate from a very high 
degree of inhomogeneity of bosonic insulators 
which is hard for a standard 
perturbative replica-trick-based approach (either mean field or one-loop expansion) to 
capture. When all low energy one-particle states are localized, 
the instability of a conventional ideal boson condensate (with a given 
number density in an infinity system) towards a local condensate with 
infinite number density appears to be evident.  
It is the finite repulsive interaction which prevents such 
collapsing 
from taking place.
This distinctly 
differs from the problem of fermion localization: 
various fixed points in the one-parameter scaling theory for 
noninteracting system seem to be stable in weakly interacting 
systems\cite{Abrahams79}.

In this Letter, I would like to 
study the renormalized chemical potential and population of bosons at
different localized states 
to illustrate the microscopic structure of bosonic insulators. 
Secondly, I will argue that interactions between bosons result in
an emergent geometry of bosonic insulators. 
This observation suggests that certain superfluid-insulator quantum 
phase transitions should be predominantly 
driven by geometrically percolating localized states.
I also apply the mean field solutions to probe the low energy Hilbert subspace
relevant to quantum critical points. 
At last I propose to study critical behaviors of these quantum phase 
transitions using an effective Hamiltonian defined in
percolating clusters.

Consider bosons 
in a d-dimension space with simple {\rm short range random} impurity 
potentials.  
All relevant low energy one-particle states are already localized (and  
are far away from the mobility edge which separates extended one-particle 
states and localized states).
Furthermore, bosons 
interact via a weak two-body 
delta-like interaction $U_0\delta({\bf r}-{\bf r}')$.
The set of wave functions I am going to exam can be written in the following form

\begin{eqnarray}
\Psi \sim \sum_{\{ n_k \}} g(\{ n_k \})
\prod_k \frac{(b_k^\dagger e^{i\phi_k})^{n_k}}{\sqrt{n_k!}} |vac>;
\label{wf}
\end{eqnarray}
$b_k^\dagger$ is the creation operator of a boson at a localized one-particle state $k$ 
of energy $\epsilon_k$ and $\phi_k$ is the corresponding phase. 
Different states correspond to
different choices of $g(\{ n_k\})$.
For an insulator in the {\em weakly} interacting limit (i.e. the dimensionless 
interaction constant $\Gamma$ is small), 
$g(\{n_k\})$ is nonzero only for a given set of integer values of 
$n_k=Int[n(\epsilon_k)]$
and no resonance is
involved. $n(\epsilon)$ as a function of $\epsilon$ 
is calculated below and depends on both interactions and disorder 
strength.
Here $Int[I+\eta]$ takes integer value $I$ if $1/2 >\eta>0$
and $I+1$ if $1 > \eta >1/2$; at $\eta=1/2$ it is equal to either
$I$ or $I+1/2$. 
For a superfluid phase, $g$ is a smooth function within a wide range of distribution
$\{n_k\}$; an extreme case is the equal-amplitude state 
with $g(\{n_k\})$ being constant.

Consider an insulator in the weakly interacting limit
where the spatial overlap between different occupied states is
negligible; 
$g(\{ n_k \})=\prod_k\delta_{n_k, Int[n(\epsilon_k)]}$. 
The total (mean field) energy for such a state which 
is fully characterized 
by
$\{ n(\epsilon_k) \}$ and is independent of $\{ \phi_k\}$ (see 
Eq.(\ref{wf}) and discussions following it)
can be shown to be

\begin{eqnarray}
&& E_0(\{ n(\epsilon_k)\}, \{ \phi_k\})=\int d\epsilon (\epsilon-\mu) 
\nu(\epsilon) n(\epsilon)
\nonumber \\
&& + U_0  {\int} d\epsilon d\epsilon' n(\epsilon)n(\epsilon') 
C(\epsilon,\epsilon'),
\\ \nonumber
&& C(\epsilon,\epsilon')=\int d^d{\bf r} \nu(\epsilon, {\bf r}) 
\nu(\epsilon', {\bf r}).
\label{Energy}
\end{eqnarray}
$\mu$ is the chemical potential to be determined self-consistently.
$\nu(\epsilon)$ and $\nu(\epsilon, \bf r)$ are the density of states 
and local density of states respectively,
$\nu(\epsilon, {\bf r})$ $=\sum_k \delta(\epsilon-\epsilon_k)\psi^*_k({\bf 
r})\psi_k ({\bf r})$,
$\nu(\epsilon)=\int d^d{\bf r} \nu(\epsilon, \bf r)$.
$k$ is a quantum number labeling one-particle eigenstates in the disordered system.
Finally, the distribution is subject to a constraint that the 
average number density is $\rho_0$, i.e.,
$\int d\epsilon n(\epsilon) \nu(\epsilon)$$=V_0\rho_0$.
($V_0$ is the Volume.) Minimizing  
$E_0(\{n(\epsilon_k)\})$ with respect to $\{ n(\epsilon_k)\}$ suggests the following 
integral equation for the ground state
population distribution ($\epsilon < \mu$)

\begin{equation}
2 U_0\int d\epsilon' n(\epsilon') C(\epsilon,\epsilon') 
=\nu(\epsilon)(\mu-\epsilon).
\label{distribution}
\end{equation}

I practically seek for a solution when 
interactions are weak and the chemical potential $\mu$ is much smaller than 
$\Delta_0=( \nu(\mu) \xi_L^d)^{-1}$, the {\em typical} single particle 
level spacing at chemical potential in a volume defined by the 
localization 
length $\xi_L$
(And for simplicity, we 
neglect the energy dependence of $\xi_L$
for all states under consideration are far away from 
the mobility edge.).
In the leading order of $\mu/\Delta_0$,
I obtain the solution to Eq.(\ref{distribution})(after an average over random potentials)
in d-dimension\cite{Zhou04}

\begin{eqnarray}
&& n(\epsilon)=\frac{N_L}{2\Gamma}\frac{\mu-\epsilon}{\epsilon_F},
\frac{\mu}{\epsilon_F}=(\frac{\Gamma}{N_L})^{{2}/{(d+2)}}S
\label{average}
\end{eqnarray}
when $\epsilon \leq \mu$;
otherwise, $n(\epsilon >\mu)=0$.
Two dimensionless parameters entering the solution 
are
\begin{equation}
N_L=\xi_L^d \rho_0, \Gamma=\frac{U_0\rho_0}{\epsilon_F}
\end{equation}
which characterize the strength of disorder and interactions respectively.
$\epsilon_F=\rho_0^{2/d}/2m$ is the Fermi energy for free fermions with the identical mass 
($m$) and density ($\rho_0$);
$S$ is a constant of order of unity (I have set $\hbar=1$). 
I also assume that $N_L$ is much larger than one.
It is worth remarking that $\mu$, the {\em renormalized} chemical potential of disordered bosons
is proportional to $\rho_0^{2/(d+2)}$.

The microscopic properties of the insulating states are fully characterized by this
distribution function. For instance, one can evaluate $\xi_T$, the typical 
distance between
two states occupied by weakly interacting bosons, the typical 
number 
density of 
bosons
in each occupied state $\rho_L$ and the ratio between $\Delta_0$ and $\mu$.
I list the results here

\begin{eqnarray}
\frac{\Delta_0}{\mu}\sim
\frac{\rho_L}{\rho_0}= 
(\frac{\xi_T}{\xi_L})^d=
\frac{1}{N_L}(\frac{N_L}{\Gamma})^{d/(d+2)}
\label{parameters}
\end{eqnarray}
which have been expressed in terms of $N_L$ and $\Gamma$.
Following Eq.({\ref{parameters}) that in a dilute limit  
where
$\xi_T$ is much larger than $\xi_L$, or
$\Gamma \ll N_L^{-2/d} (\ll 1)$,
the local level spacing $\Delta_0$ is 
much larger than $\mu$.

If the interaction is indeed very weak so that $\xi_T \gg 
\xi_L$, then typical
finite-size condensates at various localized states with the density 
$\rho_L$ are spatially 
disconnected. 
For a typical localized state $k$,
the number of particles in the ground state is unique except when
$n(\epsilon_k)$ is a half-integer.
The energy gap in local particle-like or hole-like excitation spectra  
can be shown to be proportional to 
$\{ 1 \pm 2 Int[n(\epsilon_k)]\mp 2n(\epsilon_k)\}$ $\Gamma \mu_F/N_L$
and distribute
in a range between $0$ to $a_0 \Gamma \mu_F /N_L$ ($a_0$ is a quantity of order of 
unity).
One therefore obtains the typical on-site temporal correlation function 
$G^T(\tau,0)=<b^+(\tau,{\bf r})b(0,{\bf r})>$,

\begin{equation}
G^{T}(\tau,0)=\exp(-\frac{\tau}{\tau_c}), \tau_c \sim
\frac{N_L}{\Gamma \epsilon_F}
\label{typical}
\end{equation}
which reflects the well-known $U(1)$-symmetry restoring
occurring {\em independently} at different localized states. 
(the superscript $T$ indicates a typical value).
For the non-interacting case, 
$\tau_c$ goes to infinity as it should for ideal bosons.

When $n(\epsilon_k)$ is precisely a half integer, however lowest energy
states are two-fold
degenerate. 
These states play an 
important role in the calculation of
the averaged Green's function and can yield to an ensemble 
averaged power-law temporal 
correlation and divergent
susceptibility.

Eq.(\ref{average}) indicates that
locally, {\em many-body} collective states with $Int[n(\epsilon_k)]$ 
$Int[n(\epsilon_k)]\pm 1$-, $Int[n(\epsilon_k)]\pm 2$-, 
... bosons form a low energy Hilbert subspace.
To establish a phase coherence and superfluidity,
a local condensate $k$ with $n_k=Int[n(\epsilon_k)]$ 
bosons has to {\em resonate} with low lying states 
of $n_k\pm 1$-, $n_k\pm 2$-, 
... bosons.
For 
studies of the physics close to insulator-Superfluid quantum 
phase 
transitions, one has to therefore take into account all the low energy states
represented by 
$n_k=Int[n(\epsilon_k)]$, $Int[n(\epsilon_k)]\pm 1$,
$Int[n(\epsilon_k)]\pm 2$ etc.
A convenient basis to discuss resonance between these states is the 
following 
coherent state 
representation

\begin{eqnarray}
&&
|\{\phi_k\}>\approx \prod_{k} \sum_{n_k} g_0(n_k -Int[n(\epsilon_k)]) 
 \frac{(b_k^\dagger e^{i\phi_k})^{n_k}}{\sqrt{n_k!}} |vac>,
\nonumber\\
&& <\{\phi_k\}|\{\phi'_k\}>\approx\prod_k \delta(\phi_k-\phi'_k).
\label{csr}
\end{eqnarray}
$g_0(n)$ is unity for $n_{max} > |n| >0$; 
$n_{max}$ is much larger than one but smaller
than $n(\epsilon_k)$ so that {\em all the low energy states to be 
involved in resonance are present } in this representation.

The energy of these states, in addition to that in Eq.(\ref{Energy})
acquires contributions from the the exchange (random) coupling 
between two resonating condensates. 
This consideration and Eqs.(\ref{Energy}),
(\ref{typical}) indicate that
close to insulator-superfluid transitions,
the effective Hamiltonian in the {\em truncated low energy subspace} 
spanned by 
states
in Eq.(\ref{csr}) should be\cite{Zhou04}

\begin{eqnarray}
&& {\cal H}_{eff}= \sum_{all k} E_c(k) (\hat{n}_k-n(\epsilon_k))^2
\nonumber \\
&& + \sum_{\cal C}
\sum_{<kl> \in {\cal C}}
J_{kl} \cos(\phi_k-\phi_l);\nonumber \\ 
&& E_c=\frac{B_k}{\tau_c}, J_{kl}= \frac{(n(\epsilon_k)n(\epsilon_l))^{1/2}}{\rho_L\xi_L^d} A_{kl} J.
\label{pH}
\end{eqnarray}
Here $\hat{n}_k=b^\dagger_k b_k$ is the number operator of
bosons at state $k$; and $[\hat{n}_k,\phi_l]=-i\delta_{k,k'}$.
The magnitude of ${n}(\epsilon_k)$ 
is given in Eq.(\ref{average}).
The amplitude of $J_{kl}$ is given by $J=\rho_L \xi_L^d \mu$;
$B_k \sim$ $\xi_L^d \int d{\bf r} |\psi_k^*({\bf 
r})\psi_k({\bf r})|^2$, 
and the overlap integral between two localized states 
is $A_{kl}\sim$ 
$\int \psi^*_k({\bf r}) \psi_l({\bf r})$ 
${\rho({\bf r})}/{\rho_L} d{\bf r}$; $\rho({\bf r})$ is the local boson density.
Finally, in deriving Eq.(\ref{pH})
I have 
neglected a) exchange interactions between pairs of states  
of distance much longer than $\xi_L$ because $A_{kl}$,
the overlap integrals of those pairs are exponentially small;
b) exchange couplings involving four contacting localized condensates.
Only exchange couplings between two contacting localized 
states 
separated by distance smaller than $\xi_L$ (labeled as $<kl>$ in the 
second sum) are kept.
Configurations of contacting localized states define clusters 
$\{\cal C\}$ 
of various sizes and represent {\em 
resonating} localized condensates.
Note that 
for a given $\Gamma$, properties 
of an insulating state are determined by the Hamiltonian in Eq.(\ref{pH}) 
defined 
in 
percolating clusters $\{{\cal C} \}$ of resonating states;
the distribution of clusters $\{\cal C \}$ 
itself is a function of $\Gamma$ as an emergent property of insulating 
states.

The ratio between the above two energies, $1/\tau_c$ and $J$ is

\begin{equation}
\frac{1}{J\tau_c}=a (\frac{\Gamma}{N_L})^{2d/(d+2)},
\label{ratio}
\end{equation}
which is much less than unity when $\Gamma \ll N_L$. 
($a$ and $b$ introduced below are constants of order of unity.)
Close to the percolation threshold discussed below,
$1/J\tau_c$ is proportional to $1/N_L^2$ and turns out to be much smaller than 
unity as $N_L$ in our case is much less than one.

In a {\em mean field approximation} where $1/J\tau_c$ is set to be zero,
phases of two localized condensates are locked once 
they are in contact. 
$G^T_{\cal C}(\tau,0)$,
the typical temporal Green's function 
of bosons belonging to a given cluster ${\cal C}$ (with random chemical 
potentials of course), is a 
function of the size $s$ of the cluster ${\cal C}$ only.
For a given cluster, this Green's function is still given by
Eq.(\ref{typical}); however, $\tau_c$ is replaced with $s \tau_c$ 
implying the slow down of the quantum
symmetry restoring in large clusters.

Solutions to Eq.(\ref{pH}) can be obtained in this mean field 
approximation ($d>2$ for the rest of discussions). When all clusters $\{ 
C_\infty \}$ present are infinite, 
the following wave function of a {\em resonating} 
state with 
various clusters represents one of such solutions\cite{Zhou04},

\begin{eqnarray}
&& |g> \approx \prod_{{\cal C}_\infty} \prod_{k \in {\cal C}_\infty} \sum_{n_k} 
g_0(n_k-Int[n(\epsilon_k)])
\frac{(b_k^\dagger e^{i\phi_k})^{n_k}}{\sqrt{n_k!}}  
\nonumber \\
&& \prod_{l \neq k } \sum_{\{n_l\}} \delta_{n_l, Int[n(\epsilon_l)]}
\prod_l \frac{(b_l^\dagger )^{n_l}}{\sqrt{n_l!}} |vac>
\label{cluster}
\end{eqnarray}
where the second term corresponds to the wave function of localized bosons that do 
not belong to any clusters ${\cal C}_\infty$. The set of variables
$\{\phi_k, k \in {\cal C}_\infty \}$ minimize the exchange couplings in Eq.(\ref{pH}). 
Again $g_0(n_k-Int[n(\epsilon_k)])$ is a smooth function in the vicinity of 
$n_k=Int[n(\epsilon_k)]$ which stands for resonance between localized 
condensates with different 
occupation numbers.

Close to the percolation threshold,
one obtains the following formula of the Green's function

\begin{eqnarray}
&& G^T(\tau,0)=\sum_s P(s) \exp(-\frac{\tau}{s \tau_c});
\label{typical1}
\end{eqnarray}
$P(s)$ is the probability of finding a localized state belonging to a 
cluster of size $s$ and exhibits 
interesting scaling behaviors (for general discussions, see 
Ref.\onlinecite{Plischke94,Chaikin00}). 
Let us define that the concentration of occupied states as
$p(\Gamma, N_L)$ $=b ({\xi_L}/{\xi_T})^d$ $=b 
{N_L}({\Gamma}/{N_L})^{d/(d+2)}$.
The one-parameter scaling ansatz suggests that
$P(s)\sim P_0(s/\xi(p))$. Thus, one arrives at
the following scaling form
$ G^T(\tau,0)\sim 
g_t({\tau}/{\tau_c \xi(p)})$;
$g_{t}(x)$ approaches zero when $x$ goes to infinity.
Here $\xi(p)$ is the percolation correlation length expected to diverge as
$|p-p_c|^{-\nu}$;
close to the critical interacting constant $\Gamma^L_c$,
one generally has $\xi(\Gamma, N_L)\sim (\Gamma^L_c-\Gamma)^{-\nu}$.

Eq.(\ref{typical1}) and discussions above therefore indicate that 
$U(1)$-symmetry breaking 
should take 
place if and only if the percolating threshold is reached so that $\xi(p)$ 
is infinite. 
The appearance of infinitely percolating clusters (with a finite density) 
turns out to be necessary and sufficient (in a mean field approximation)
for the development of long range correlations. 
An estimate based on Eq.(\ref{parameters}) yields the critical 
point for geometric quantum phase transitions 

\begin{equation}
\Gamma^L_c = (\frac{p_c}{b})^{-d/(d+2)} N_L^{-2/d}.
\label{lower}
\end{equation}
The superscript $L$ stands for a lower critical point for the 
superfluid phase.
As $d$ approaches infinity, 
the percolation threshold $p_c$ scales as
$1/d$ and clusters of infinite size 
first appear in the dilute limit where $\xi_T \gg \xi_L$.
The percolation can be studied in a mean field
approximation\cite{repulsion}.

Furthermore,
the superfluid density above the percolation threshold $p_c$ can also be
related to clusters extending to infinity. 
One can easily establish that 
$\rho^U_{s}$, the upper bound of the superfluid density $\rho_s$
should be proportional 
to $n_\infty(\Gamma)$, the probability density for a local state to be in 
an infinite cluster appearing above $\Gamma^L_c$.
$\rho^U_s
= \rho_0 n_\infty 
(\Gamma) \sim (\Gamma-\Gamma^L_c)^\beta$.
In the mean field approximation, the critical exponents are $\nu=1/2$, $\beta=1$.

The above mean field approximation appears to be oversimplified from the point of
view of Shklovskii-De Gennes backbone (SDeGb) model for percolating 
clusters\cite{Skal75,DeGennes76}.
In fact, the SDeGb model indicates that percolating clusters consist of dense 
regions connected
by one-dimensional strands even when $d>1$. Early results further imply that
it should be unlikely for a 1D infinite quantum system to be 
ordered\cite{Hartman95}.  
So at any small but finite value of $1/J\tau_c$, we expect, after taking into account 
quantum effects, that the onset superfluidity appears
above the percolation threshold. Particularly, 
the true quantum transition takes place when bosons along 1D strands (with finite 
length) in the SDeGb model become correlated in phases.
Thus, the lower critical value $\Gamma_{cQ}^L$
receives a quantum correction,

\begin{equation}
\frac{\Gamma^L_{cQ}- 
\Gamma_c^L}{\Gamma_c^L}\sim (\frac{\xi_L}{\xi_{1d}(1/J\tau_c)})^{1/\gamma}.
\label{qcv}
\end{equation}
Here $\xi_{1d}(\alpha)$ is the correlation length of bosons  
along 1D strands in a Shklovskii-de Gennes cluster and is expected to be 
exponentially long compared to $\xi_L$ when $\alpha$ approachs zero.
$\gamma$ is the critical exponent of percolation correlation length above the threshold.
As indicated in Eq.(\ref{qcv}), the quantum correction to the lower 
critical point
$\Gamma_{cQ}^L -\Gamma^L_c$ is negligible when
$1/J\tau_c$ is much smaller than unity.
Though in this case the critical phenomenon might differ from the 
percolative ones, the size of the critical regime where the deviation from the mean 
field critical behavior
becomes 
evident could be {\em exponentially} small as Eq.(\ref{qcv}) which sets this size
implies.

At last let us consider what happens when interactions are further 
increased 
and $\Gamma \gg N_L$.
Bosons at different localized states then prefer to have independent 
dynamics; that is, 
$G^T_{\cal C}(\tau,0;0,0)\sim \exp(-{\tau}/{\tau_c})$ 
and therefore $G^T(\tau,0)$ discussed in Eq.(\ref{typical1}) remains short 
ranged 
even when 
there are infinite clusters\cite{unbound}.
The above argument suggests that the upper critical point for the superfluid 
phase should be set by the following equation:
$\Gamma^U_c=c N_L$ and $c$ is a number of order of unity.
The superfluid phase appears when $\Gamma_c^U > \Gamma > \Gamma_c^L$.

When $N_L$ is much less than one as in the presence of strong disorder,
$\Gamma^U_c \ll \Gamma_c^L$ and the regime where the superfluid 
phase exists shrinks to zero. 
Indeed, 
bosons in each localized state have independent dynamics,
either because of the absence of connected clusters ($\Gamma \ll 
\Gamma^L_c$) 
or when above the percolation threshold,
because of the rapid quantum symmetry restoring ($\Gamma >\Gamma^L_c 
\gg \Gamma_c^U$). 
No insulator-superfluid phase transitions are likely to occur in this 
case and there is one single insulating phase\cite{unbound}.

To summarize, I show that 
bosonic insulators of interacting bosons have fascinating geometric properties
which are characterized by the parameters in Eq.(\ref{parameters}).
Certain quantum phase transitions from insulators to superfluid phases
as the interaction constant varies
are strongly influenced by the emergent geometries of insulating states.
This work is in part supported by NSERC, Canada and a grant from UBC. 
I would like to thank P. W. Anderson, B. Bergersen, 
G. Refael, S. Sachdev, B.Spivak 
and P. B. Wiegmann for helpful discussions.


\begin{thebibliography}{99}
\bibitem{repulsion} 
The effect of level repulsion on 
percolation can be taken into account by redefining
the concentration $p$.
\bibitem{unbound} The impurity potentials considered in this article are unbounded;
for bounded potentials as those introduced and studied in 
Ref.\onlinecite{Fisher89,Scalettar91,Krauth91}, there 
can be more than one insulating phases.
\bibitem{coldatom}For cold atoms, the dimensionless coupling constant 
$\Gamma$ can be experimentally varied either
by varying the two-body scattering length close to the Feshbach 
resonance or simply varying the density $\rho_0$.
\end{thebibliography}
\end{document}